\documentclass[aps,prb,showpacs,twocolumn,superscriptaddress]{revtex4}

\usepackage{amsmath}
\usepackage{graphicx}
\begin{document}

\title{Evaluation of Reverse Monte Carlo Models based on Molecular
  Dynamics Simulations: A Case Study of Ion Conducting Network Glasses}

\author{Christian R.\ M\"uller}
\affiliation{Institut f\"ur Physik, Technische Universit\"at
  Ilmenau, 98684 Ilmenau, Germany}
\author{Vindu Kathriarachchi}
\affiliation{Department of Physics, Central Michigan University, Mount Pleasant, Michigan 48859, USA}
\author{Michael Schuch}
\affiliation{Institut f\"ur Physik, Technische Universit\"at
  Ilmenau, 98684 Ilmenau, Germany}
\author{Philipp Maass}
\affiliation{Fachbereich Physik, Universit\"at
  Osnabr\"uck, Barbarastra{\ss}e 7, 49069 Osnabr\"uck, Germany}
\email{philipp.maass@uni-osnabrueck.de}
\homepage{http://www.statphys.uni-osnabrueck.de}
\author{Valeri G.\ Petkov} 
\affiliation{Department of Physics, Central Michigan University, Mount Pleasant, Michigan 48859, USA}

\date{\today}



\begin{abstract}

We investigate the quality of structural models generated by the
Reverse Monte Carlo (RMC) method in a typical application to amorphous
systems. To this end we calculate surrogate diffraction data from a
Li$_2$O-SiO$_2$ molecular dynamics (MD) simulation and use the total
scattering function, in addition to minimal pair distances and
coordination numbers of silicon (oxygen) to oxygen (silicon) ions, as
input for the RMC modeling. Then we compare partial radial
distribution functions, coordination numbers, bond angles, and ring
sizes predicted by the RMC models with those of the MD system. It is
found that partial distributions functions and properties on small
lengths scales, as distributions of coordination numbers and bond
angles, are well reproduced by the RMC modeling. Properties in the
medium-range order regime are, however, not well captured, as is
demonstrated by comparison of ring size distributions. Due care
therefore has to be exercised when extracting structural features from
RMC models in this medium-range order regime. In particular we show
that the occurrence of such features can be a mere consequence of the
chosen starting configuration.

\end{abstract}

\maketitle

\section{Introduction}
The Reverse Monte Carlo (RMC) method is commonly used to build
structure models based on experimental data. Introduced by McGreevy
and Pusztai in 1988,\cite{McGreevy/Pusztai:1988} it has been spreading
fast and is now considered a standard method in analyzing structural
data. Advantages of this method are its easy implementation and its
wide applicability. It has been used to model various material systems
such as crystals, polymers and glasses. In principal any structural
data can be used as input for the RMC method, but most modelings focus
on using diffraction data obtained from X-ray and/or neutron
scattering. In ion conducting glass systems RMC models of the
structure of the structure have been created for $x$Li$_2$S+$(1-x)$
SiO$_2$ glasses,\cite{Uhlig/etal:1996} 0.7SiO$_2$+0.3Na$_2$O
glass,\cite{Fabian/etal:2007} $x$Na$_2$S+$(1-x)$B$_2$S$_3$
glasses\cite{Yao/etal:2005} and 0.5Li$_2$S+0.5[$(1-
  x)$GeS$_2$+$x$GeO$_2$] 
glasses\cite{Messurier/etal:2009} among others.

As pointed out by McGreevy, RMC models are ``neither unique nor
'correct' {''}, but can aid our understanding of local structure
properties and their relation to other physical
properties.\cite{McGreevy:2001} Accordingly it is important to qualify
RMC models for different material classes and to get insight into the
limits of this method. This is becoming a more urgent question now,
since in recent RMC studies not only the short range order of various
network glasses has been investigated, but also the medium range
order. Among those studies are discussions of the rings sizes in
vitreous SiO$_2$ and GeO$_2$,\cite{Kohara/Suzuya:2005} a detailed
investigations of amorphous GeSe$_2$,\cite{Murakami/etal:2007} and a
proposal of a structural model for multi-component borosilicate
glasses, where partial segregation of silicon and boron rich regions
is predicted.\cite{Fabian/etal:2008} It was also suggested to use such
models as basis for further investigation of possible conduction
pathways of the mobile ions. In this respect the RMC models have been
employed in connection with geometric constraints and the Bond Valence
(BV) analysis\cite{Adams/Swenson:2000,Swenson/Adams:2003} (see Ref.
\onlinecite{Mueller/etal:2007} for a critical discussion of this
procedure).

In this paper we test the RMC method against structural data obtained
from a Molecular Dynamics (MD) simulation of a Li$_2$O-SiO$_2$ glass.
For this purpose we calculate surrogate diffraction data from the
simulated MD structures and these surrogate data are used as input for
the RMC modeling. For the evaluation of the resulting RMC models we
determine how well various properties of the original MD structure are
reproduced. Particularly we compare properties such as partial radial
distribution functions and ring-size distributions, which are not
easily accessible by experiment. Through our evaluation it can be
clarified how far one can use the RMC method to gain insight into
these properties, and where one has to be cautious to take features of
the RMC model for real.

We want to stress that for the testing to be valid, it is not
necessary that the MD simulation is a particularly good representation
of the real lithium silicate glass. Rather, the MD structure can be
seen as a valid glass system in itself. We will show that the RMC
models generally compare well with the MD structure, but that one has
to take care when analyzing features of the medium-range order regime.

\section{\label{sec:MD-setup} Molecular Dynamics Simulations}
We perform MD simulations of a lithium silicate glass with the
chemical formula Li$_2$O-SiO$_2$ using the potential model of
J.~Habasaki and I.~Okada.\cite{Habasaki/Okada:1992} The cubic
simulation box has a length $L=50.04$~{\AA} \space and contains 11664
atoms (3888 Li, 1944 Si, 5832 O) corresponding to a density of
$2.27$~g/cm$^3$ and a number density of $\rho_0=0.093$~{\AA}$^{-3}$.
Periodic boundary conditions are used. The simulations are performed
in the NVE ensemble (micro canonical ensemble where the number $N$ of
particles, the volume $V$ of the simulation box, and the total energy
$E$ are kept constant). The energy $E$ was adjusted so that the
temperature of the system fluctuates around a mean value of 301~K with
deviations of 2~K. The systems are equilibrated for about 1~ns and the
runs for obtaining data have a duration of 2~ns, using a time step
interval of $\Delta t=1$~fs.

The effective interatomic interactions between two atoms of type $i$ and $j$ at distance $r$ are:
\begin{equation}
U_{ij}(r)= \frac{e^2}{4\pi\varepsilon_0} \frac{z_i z_j}{r}+ f_0 (b_i
+b_j) \exp \left( \frac{a_i+a_j-r}{b_i+b_j} \right)
-\frac{c_i c_j}{r^6}
\label{eq:Uij}
\end{equation}
where the parameters listed in Table~\ref{tab:params} have been
optimized \cite{Habasaki/Okada:1992} and shown to give good agreement
with experimental data
\cite{Habasaki/Okada:1992,Habasaki/etal:1995,Banhatti/Heuer:2001,Heuer/etal:2002}.
The interaction potential in Eq.~(\ref{eq:Uij}) is composed of three
terms. The first one in (\ref{eq:Uij}) is the Coulomb interaction with
effective charge numbers for the species. The second term is a
Born-Meyer type potential, which takes the short-range repulsive
interactions into account, and the third is a dispersive van-der-Waals
interaction. It is only used for interactions involving oxygen.

\begin{table}
\caption{\label{tab:params} Potential parameters for the MD-simulations
(cf.\ Eq.~(\ref{eq:Uij})).}
\begin{center}
\begin{tabular}
{c|@{\hspace{1em}}c@{\hspace{1em}}@{\hspace{1em}}c@{\hspace{1em}}@{\hspace{1em}}c@{\hspace{1em}}@{\hspace{1em}}c@{\hspace{1em}}}\hline\hline
 Ion & $z$ & $a$ [{\AA}] & $b$ [{\AA}] & $c$ [{\AA}$^3\sqrt{\rm kJ/mol}$] \\
 \hline  Li$^+$ & \phantom{-}0.87 & 1.0155 & 0.07321 &
 \phantom{1}22.24 \\  Si$^{4+}$ & \phantom{-}2.40 & 0.8688 &
 0.03285 & \phantom{1}47.43 \\  O$^{2-}$ & -1.38 & 2.0474 &
 0.17566 & 143.98 \\ \hline
 \multicolumn{5}{c}{$f_0=4.186$\,kJ{\AA}$^{-1}$mol$^{-1}$
\hspace{1em} $r_{\rm c}=1.3$\,{\AA}}\\ \hline\hline
\end{tabular}
\end{center}
\end{table}

The system was prepared by putting the atoms on a cubic crystal
lattice and assigning to every atom random velocities drawn from a
Maxwell-Boltzmann distribution corresponding to a temperature of
2500~K, which is well above the (computer) glass transition
temperature of this system. From this liquid state the system is
cooled down in several steps with intermediate periods of
equilibration. First an NVT run (canonical ensemble, where the number
$N$ of particles, the volume $V$ and the temperature $T$ are fixed) of
10~ps at 2500~K is performed, followed by an NVE run of the same
duration. After simulating another 20~ps in the NVT ensemble and 10~ps
under NVE conditions the temperature is decreased in four subsequent
sequences down to 300~K. Each cooling cycle consists of a 10~ps run
using a thermostat to decrease the temperature linearly, a 10~ps NVT
run at the target temperature, and a 10~ps NVE run to verify that
there are no temperature drifts. The configurations at the end of the
300~K cooling cycle are used as starting points for a 800~ps long
equilibration run using the NVE ensemble. The measuring runs are 2~ns
long. All MD-simulations were carried out with the LAMMPS software
package \cite{LAMMPS}.

The partial and total radial distribution functions $g_{ij}(r)$ and
$G^{\rm PDF}(r)$ as well as the total scattering structure factor
$S(Q)$ were calculated according to the PDF-formalism (see
\onlinecite{Keen:2000} for a discussion of different possible
definitions of scattering functions). The partial radial distribution
functions are given as
\begin{equation}
\label{eq:rdf_part}
g_{ij}(r)=\frac{n_{ij}(r)}{4\rho_j \pi r^2{\rm d}r} \hspace{.25 cm},
\end{equation}
where $n_{ij}(r)$ is the average number of particles of type $j$
between distances $r-{\rm d}r/2$ and $r+{\rm d}r/2$ from a particle of
type $i$, and $\rho_j$ is the mean number density of particles of type
$j$.

The total radial distribution function is calculated by
\begin{equation}
\label{eq:rdf_tot}
G^{\rm PDF}=4\pi r\rho_0\left[\sum_{i,j=1}^m(w_{ij}g_{ij})-1\right]
\hspace{.25cm},
\end{equation}
where $\rho_0$ is the total number density of the system, $m$ is the
number of particle types, and $w_{ij}$ are weighting factors:
\begin{equation}
\label{eq:rdf_weight}
w_{ij}=\left(\sum_{i=1}^mc_i\bar{b}_i\right)^{-2}\sum_{i,j=1}^mc_ic_j\bar{b}_i\bar{b}_j \hspace{0.25cm}.
\end{equation}
Here $c_i=\rho_i/\rho_0$ are the molar fractions of particles of type
$i$, and $\bar{b}_i$ is their average bound coherent scattering
length. In order to calculate X-ray diffraction functions one has to
replace the $\bar{b}_i$ with the atomic form factors $f_i$.

Finally, the total structure factor $S(Q)$ is calculated from $G^{\rm PDF}(r)$ by
\begin{equation}
\label{eq:sq}
Q[S(Q)-1]=\int\limits_{0}^{\infty} G^{\rm PDF}(r)\sin Qr {\rm d}r \hspace{.25cm}.
\end{equation}

\begin{table}[t!]
\caption{\label{tab:bf} Average bound coherent 
scattering lengths and atomic form factor for Li, Si, and O.}
\begin{center}
\begin{tabular}{@{\hspace{1em}}c@{\hspace{1em}}|@{\hspace{1em}}c@{\hspace{1em}}c@{\hspace{1em}}c@{\hspace{1em}}}
\hline\hline
    & Li & Si & O \\
$\bar{b}$ & -1.9 & 4.1491 & 5.803 \\  
$f$ & 3.005 & 14.41 & 8.144 \\
\hline\hline
\end{tabular}
\end{center}
\end{table}

All data from the MD system was averaged over 11 configurations from
the 2~ns measurement run, which are 200~ps apart each. The scattering
lengths and atomic form factors used in Eq.~\ref{eq:rdf_weight} were
taken from Ref.~\onlinecite{NIST-neutron} and
Ref.~\onlinecite{NIST-xray}\nocite{Chantler:1995}\nocite{Chantler:2000},
respectively, and are listed in Table \ref{tab:bf}. In the following
we take the freedom to speak about these surrogate diffraction data
simply as ``diffraction data''and ask the reader to keep in mind that
the data was not measured but calculated from Eqs.~\ref{eq:rdf_part}
to \ref{eq:sq}.

\section{Reverse Monte Carlo modeling}

\begin{figure}[bth]
\includegraphics[width=0.4\textwidth,,clip=.]{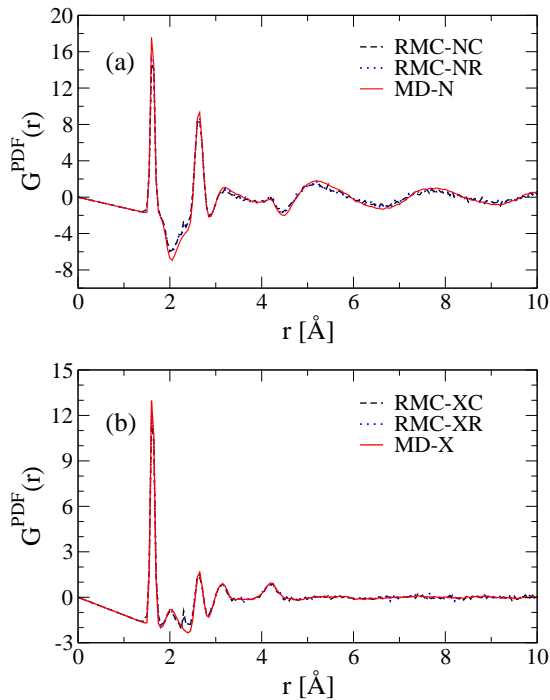}
\caption{\label{fig:Gr} (color online) Comparison of $G^{\rm PDF}(r)$
of Li$_2$O-SiO$_2$ RMC models based on crystalline (black dashed line)
and random (blue dotted line) starting configurations with that of a
Li$_2$O-SiO$_2$ MD system (red solid line). Neutron diffraction data
is shown in panel (a) and X-ray diffraction data in panel (b).}
\end{figure}

\begin{figure}[bth]
\includegraphics[width=0.4\textwidth,clip=.]{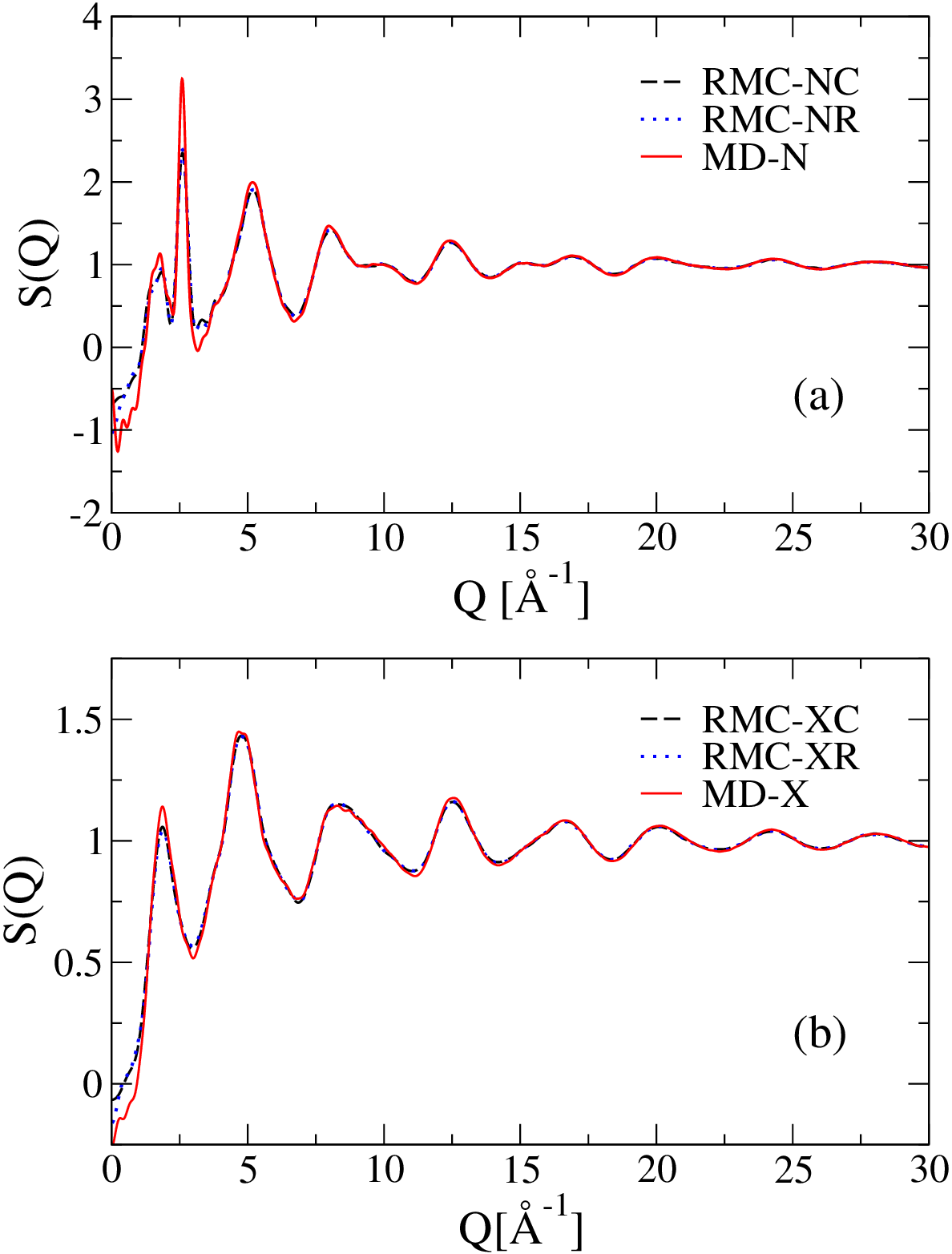}
\caption{\label{fig:Sr} (color online) Comparison of $S(Q)$ of
Li$_2$O-SiO$_2$ RMC models based on crystalline (black dashed line)
and random (blue dotted line) starting configurations with that of a
Li$_2$O-SiO$_2$ MD system (red solid line). Neutron diffraction data
is shown in panel (a) and X-ray diffraction data in panel (b). }
\end{figure}

RMC simulations were carried out using the RMC++
package.\cite{Evrard/Pusztai:2005} We started with building an initial
atomic configuration and then refined it against the $S(Q)$ and
$G^{\rm PDF}(r)$ data computed from the MD structure.

\begin{table}[b!]
\caption{\label{tab:rmin} 
Minimal atomic approach distances as used in the RMC modeling.}
\begin{center}
\begin{tabular}{@{\hspace{1em}}c|@{\hspace{1em}}c@{\hspace{1em}}c@{\hspace{1em}}c@{\hspace{1em}}c@{\hspace{1em}}c@{\hspace{1em}}c@{\hspace{1em}}}
\hline\hline
Pair & Si-Si & Si-O & Si-Li & O-O & O-Li & Li-Li \\
$r_{\rm min}$ in {\AA} & 2.8 & 1.4 & 2.5 & 2.3 & 1.7 & 2.2 \\  
\hline\hline
\end{tabular}
\end{center}
\end{table}

Two starting configurations were considered: a ``random distribution''
of atoms\cite{comm-random} (subsequently referenced by ``R'') and
another from the Li$_2$O-SiO$_2$ orthorhombic crystal structure (space
group Cmc21, subsequently referenced by ``C''). Both configurations
consists of 3000 atoms (1000 Li, 500 Si, 1500 O). The system was
chosen to be cubic with a side length of $31.82$~{\AA} so that the
atomic number density is the same as in the MD structure. {}From the
random and crystalline starting configurations, two initial models IR
and IC were prepared, respectively, by applying the following
constraints:
\begin{itemize}
\item[(i)]
Si is coordinated fourfold with O using
a minimal neighbor distance of 1.4~{{\AA}} and a maximum neighbor
distance of 1.8~{\AA}, This corresponds to a  100\% fraction $f_{\rm
  Si,O}(4)=1$ of fourfold coordinated Si.
\item[(ii)] The relative numbers of bridging (two Si neighbors) and
non-bridging oxygens (one Si neighbor) is 37\% and 60\%, respectively.
This corresponds to $f_{\rm O,Si}(2)=0.37$ and $f_{\rm O,Si}(1)=0.60$.
\item[(iii)] Minimal atomic distances $r_{\rm min}$
given in Table~\ref{tab:rmin} are required. 
\end{itemize}
Intra-tetrahedral O-Si-O angles and inter-tetrahedral Si-O-Si angles
were allowed to evolve freely. The preparation was run
until the constraints (i)-(iii) were satisfied for at least 95\% of
the atoms (for given uncertainty parameters, see below).

\begin{table}[b!]
\caption{\label{tab:rmc-sigmas} Weighting factors $\sigma_\alpha$ used for
  the input quantities in the RMC modeling.}
\begin{center}
\begin{tabular}{c|@{\hspace{1em}}c@{\hspace{1em}}c@{\hspace{1em}}c@{\hspace{1em}}c@{\hspace{1em}}c@{\hspace{1em}}c}
\hline\hline
quantity & $S(Q)$ & $G^{\rm PDF}(r)$ & $f_{\rm Si,O}(4)$  & $f_{\rm O,Si}(2)$  &
 $f_{\rm O,Si}(1)$  \\
$\sigma$ & 10$^{-6}$ &  10$^{-6}$ & 10$^{-4}$ & 10$^{-3,5}$ & 10$^{-3,5}$ \\  
\hline\hline
\end{tabular}
\end{center}
\end{table}

After creating the initial models, the final refinement is done in
order to obtain the best possible agreement between the computed
$S(Q)$ and $G^{\rm PDF}(r)$ from the RMC model and the calculated data
from the MD-simulation. Both the real space as well as the reciprocal
space data was used, since strong low-$Q$ features in $S(Q)$ emphasize
the medium range order, while $G^{\rm PDF}(r)$ shows well defined
low-$r$ features which emphasize the short range atomic order. The
same constraints as in the preparation of the initial models were
applied also during the final RMC modeling.

In the modeling, the input quantities, i.e.\ the total radial
distribution function $G^{\rm PDF}(r)$, the total structure factor
$S(Q)$, and the fractions $f_{\rm Si,O}(4)$, $f_{\rm O,Si}(2)$, and
$f_{\rm O,Si}(1)$ of differently coordinated silicon and oxygen ions
are taken into account by an effective Hamiltonian of type $H_{\rm
  eff}=\sum_\alpha \tilde\chi^2_\alpha/\sigma_\alpha^2$, where
$\tilde\chi^2_\alpha$ is the (unweighted) square deviation between the
computed and the measured (in the case of constraints presumed) value
of the input quantity $\alpha$. The weighting factors $\sigma_\alpha$
are summarized in Table~\ref{tab:rmc-sigmas}. For a detailed
description of the algorithm we refer to the manual of the RMCA and
and RMC++ package which can be downloaded at \onlinecite{rmc++-man}.

In total five RMC models were produced. Two models were generated
based on X-ray diffraction data using the random and the crystalline
starting configuration (XR model and XC model) and another two are
based on the neutron diffraction data (NR model and NC model). A fifth
model was generated using both the X-ray and the neutron diffraction
data starting from the random configuration (NXR).

In Figs.~\ref{fig:Gr} and \ref{fig:Sr} the corresponding total
structure factors and total distribution functions are compared to the
ones calculated from the MD structure. As expected, a good
agreement is achieved through the RMC modeling.

\begin{figure}[h!tb]
\includegraphics[width=0.48\textwidth,clip=.]{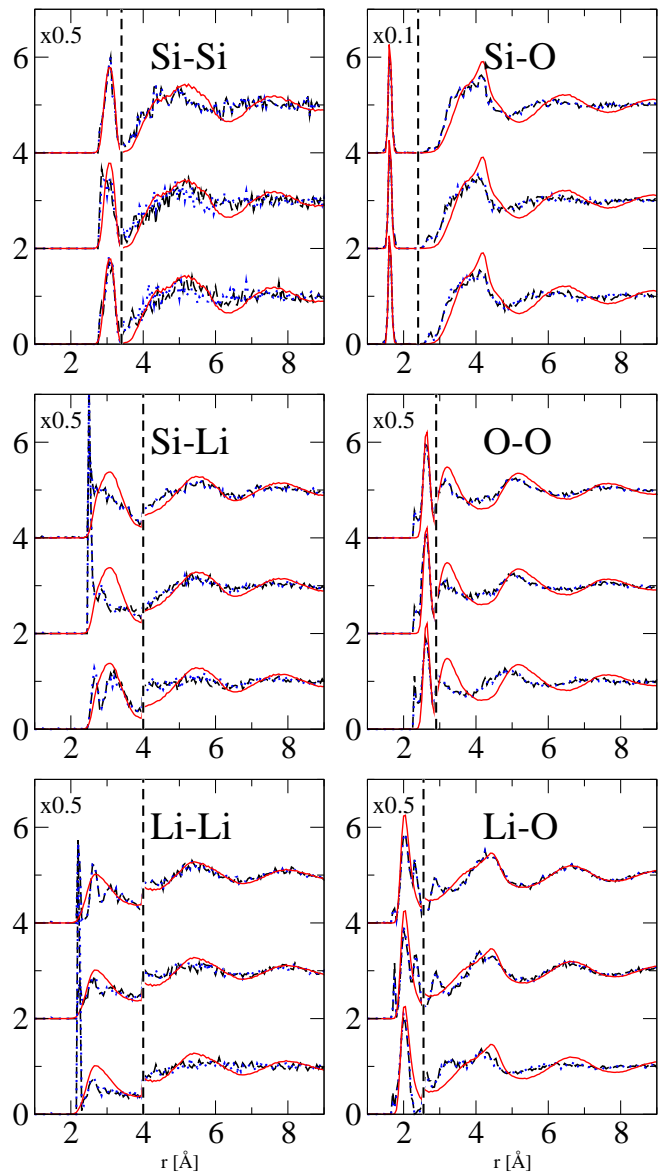}
\caption{\label{fig:prdf} (color online) Comparison of $g_{ij}(r)$ of
Li$_2$O-SiO$_2$ RMC models based on crystalline (black dashed line)
and random (blue dotted line) starting configurations with that of a
Li$_2$O-SiO$_2$ MD system (red solid line). The RMC model based on
neutron diffraction data is shown with a vertical offset of 2, while
the RMC model based on both neutron and X-ray diffraction data is
shown with an offset of 4. In order to show the full first peak, the
curves are scaled by the given factor on the left side of the vertical
dashed line in each plot.}
\end{figure}

\section{Comparison of structural properties}
\subsection{Partial radial distribution functions}\label{subsec:prdf}

In Fig.~\ref{fig:prdf} partial radial distribution functions
$g_{ij}(r)$ (see Eq.~\ref{eq:rdf_part}) are shown. Generally there is
a reasonable agreement of the RMC models with the MD structure for
Si-Si, Si-O, and Li-O. For $g_{\rm OO}$ the second peak is lacking or
too weakly pronounced in the RMC models. This difference can lead to
significant deviations of the RMC structures and the MD structure in
the medium-range order regime (see also the discussion in
Sec.~\ref{sec:rings}). While there is no dependence upon the starting
configuration (crystalline or random), some differences can be seen
between the RMC models based on the neutron diffraction data and those
based on the X-ray diffraction data. These are most pronounced in
$g_{\rm SiLi}$ and $g_{\rm LiLi}$. The RMC models for which both the
neutron and the diffraction data was used show the best agreement with
the MD data.

To summarize the result of the comparison of all RMC models and the MD
model, we calculated the integral
\begin{equation}
d_{ij}^\alpha=\int |g_{ij}^\alpha-g_{ij}^{\rm MD}|{\rm d}r
\label{eq:dij}
\end{equation}
over the difference between the partial radial distribution functions
of the RMC model $\alpha$ and the MD system. In the numerical
calculation we integrated up to $r=10$, which amounts to an
integration to infinity, since $g_{ij}^\alpha\cong g_{ij}^{\rm
MD}\cong1$ for $r\gtrsim10$. The results shown in Table~\ref{tab:prdf}
allows us to quantify the quality of the RMC models relative to each
other. It is surprising that on average the IC and IR models that do
not involve information from scattering functions, are not much worse
than the RMC models (XC, XR, NC, NR) with only one scattering probe
(X-ray or neutron). On the other hand, when information from both
scattering probes is taken into account, one obtains a significant
improvement on average. We note that this improvement is associated
with the fact that the NXR model exhibits good agreement for all
individual partial radial distribution functions. By contrast, when
the analysis is based on one scattering probe only, quite large
deviation can occur for certain partial radial distribution functions
(see, for example, $d^{\rm NC}_{\rm Si,Li}$ and $d^{\rm XC}_{\rm
Li,Li}$).

\begin{table}
\caption{\label{tab:prdf} Integrated differences $d_{ij}^\alpha$ of
partial radial distribution functions, cf.\ Eq.~(\ref{eq:dij}).}
\begin{tabular}{ccccccccc}
\hline\hline
 Pair&XC&XR&NC&NR&NXR&IC&IR&C\\\hline
 Si-Si	& 1.10 & 1.24 & 1.42 & 1.50 & 1.21 & 1.56 & 1.56 & 9.76\\
 Si-O 	& 1.25 & 1.29 & 1.44 & 1.44 & 1.28 & 3.58 & 3.47 & 6.58\\
 Si-Li	& 1.47 & 1.52 & 2.36 & 2.27 & 1.53 & 1.63 & 1.83 & 9.84\\
 O-O	& 1.44 & 1.47 & 1.03 & 0.99 & 0.88 & 2.29 & 2.21 & 5.93\\
 Li-Li	& 2.03 & 2.01 & 1.05 & 0.97 & 1.14 & 1.51 & 1.58 & 9.75\\
 Li-O 	& 1.46 & 1.42 & 1.24 & 1.12 & 0.91 & 2.01 & 2.04 & 6.71\\
\hline
 Average & 1.46 & 1.49 & 1.42 & 1.38 & 1.16 & 2.10 & 2.12 & 8.10 \\
\hline\hline
\end{tabular}
\end{table}

\subsection{Coordination numbers}\label{subsec:coord}

\begin{figure}[tbh]
\includegraphics[width=0.48\textwidth,,clip=.]{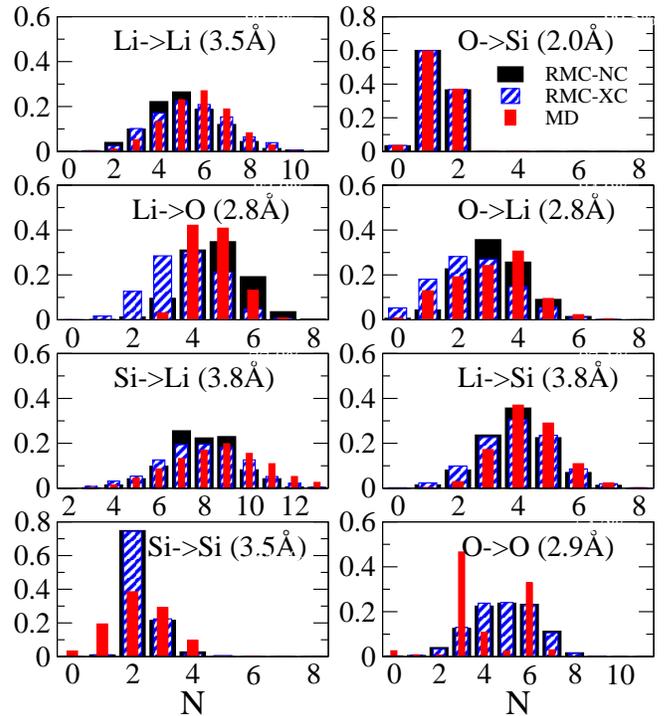}
\caption{\label{fig:coord} (color online) Comparison of histograms of
coordination numbers of Li$_2$O-SiO$_2$ for RMC models based on a
crystalline starting configuration using neutron diffraction data
(black solid bars) and X-ray diffraction data (blue striped bars), and
the original MD structure (red solid bars). The first atom type is the
center atom, and the given distances are the radii of the coordination
sphere. }
\end{figure}

The first neighbor shell coordination number distributions $f_{ij}(N)$
are defined here as the fractions of atoms of type $i$ which have $N$
atoms of type $j$ within a first neighbor shell radius $r_{ij}$. Note
that being neighbors in the sense of this analysis is not associated
with having a chemical bond. To quantify the quality of the various
RMC models $\alpha$, the overlap
\begin{equation}
\label{eq:coordoverlap}
\phi^\alpha_{ij}=1-\sum_{N=0}^\infty|f_{ij}^{\alpha}(N)-f^{\rm\scriptscriptstyle MD}_{ij}(N)|\,.
\end{equation}
was calculated. A value $\phi^\alpha_{ij}=1$ means perfect overlap between
the distributions of the coordination numbers
$f_{ij}^{\alpha}(N)$ and $n^{\rm\scriptscriptstyle
MD}_{ij}(N)$ of the RMC model $\alpha$ and the MD structure.

The results are summarized in Table~\ref{tab:coord}. No significant
differences in the $\phi^\alpha_{ij}$ are found between the RMC models
based on the crystalline starting configuration and the random
starting configuration (with one exception for Si-Li,
where a larger difference is observed between the RMC-XC and RMC-XR
models). This suggests that the quality of reproducing coordination
numbers is independent of the starting configuration. A significant
difference between the neutron based and the X-ray based RMC models is
found in the Li-O coordination numbers, where the better overlap for
the neutron based model can be traced back to the higher relative
weight of lithium in the neutron diffraction functions. An improved
overall agreement is achieved when using both X-ray and neutron
diffraction data, though the two most significant discrepancies (Si-Si
and O-O) are still there. It is also informative to take a look at the
overlap numbers of the initial models that are on the constraints only.
These are comparable in quality with the RMC models, which in addition
take into account the information from one scattering probe.
As for the partial radial distribution functions discussed in the
previous Sec.~\ref{subsec:prdf}, the RMC-NXR shows a clear improvement
compared to the initial RMC models IC and IR.

Figure~\ref{fig:coord} shows a detailed comparison of the coordination
number distributions of the RMC-NC and RMC-XC models with the MD
model. For the distribution $f_{\rm SiO}(N)$ not shown in
Fig.~\ref{fig:coord}, we obtained a very good agreement which
essentially results from the constraint that silicon atoms must have
4-fold coordination. The most striking discrepancies between the RMC
models and the MD structure are found in $f_{\rm OO}(N)$ and $f_{\rm
SiSi}(N)$. The MD model shows a clear bimodal distribution with maxima
at 3 and 6 neighbors (corresponding to non-bridging oxygens and
bridging oxygens) in $n^{\rm MD}_{OO}(N)$, while the RMC models have a
broad smooth distribution. On the other hand, $n^{\rm RMC}_{\rm
SiSi}(N)$ is much narrower than $n^{\rm MD}_{\rm SiSi}(N)$. These
findings suggest that the short-range order of the RMC models
corresponds quiet well to that of the MD structure, but that the
medium-range order, and particularly the structure of the Si-O
network, has significant differences. We note, that there are
virtually no differences between $f_{\rm SiSi}(N)$ and $f_{\rm OO}(N)$
among the five RMC models.

\begin{table}
\caption{\label{tab:coord} Overlap of coordination numbers in percent
  (see Eq.~\ref{eq:coordoverlap}); first neighbor shell radii are
  given in the second column.}
\begin{tabular}{cccccccccc}
\hline\hline
 Pair & $r_{ij}$ [{\AA}]&XC&XR&NC&NR&NXR&IC&IR&C\\\hline
 Si-O 	& 2.0	& 98.4 & 98.8 & 99.7 & 98.4 & 95.4 & 99.7 & 98.8 & 93.2\\
 O-Si 	& 2.0	& 99.5 & 99.6 & 99.5 & 99.6 & 99.1 & 99.6 & 99.6 & 91.2\\
 Li-O 	& 2.8	& 69.8 & 68.2 & 83.0 & 85.1 & 87.7 & 69.3 & 69.8 & 48.8\\
 O-Li	& 2.8	& 84.5 & 82.5 & 84.9 & 84.9 & 86.6 & 83.1 & 84.5 & 57.6\\
 Si-Li	& 3.8	& 90.4 & 80.9 & 77.8 & 78.5 & 83.8 & 80.6 & 74.5 & 47.7\\
 Li-Si	& 3.8	& 89.5 & 87.9 & 87.1 & 86.7 & 89.2 & 84.7 & 81.7 & 49.0\\
 Si-Si	& 3.5	& 64.0 & 65.1 & 64.5 & 61.9 & 62.3 & 66.0 & 64.1 & 63.9\\
 O-O	& 2.9	& 52.9 & 52.0 & 54.2 & 54.3 & 58.9 & 51.6 & 51.1 & 89.9\\
 Li-Li	& 3.5	& 82.4 & 81.5 & 79.5 & 81.2 & 92.8 & 84.4 & 84.9 & 52.6\\
\hline
Average & --	& 81.3 & 79.6 & 81.1 & 81.2 & 84.0 & 79.9 & 78.8 & 66.0\\
\hline\hline
\end{tabular}
\end{table}

In summary we can conclude that most features in the coordination
number distribution are already captured by the constraints. This may
not be surprising, since coordination numbers for Si and O have been
used as input requirements together with the rather high density of
the system. As a consequence, there is not much freedom for the
coordination numbers between other types of ion pairs.

\subsection{Bond-angle distribution}
We calculated bond-angle distributions for intra-tetrahedral angles
(O-Si-O) and inter-tetrahedral angles (Si-O-Si) and found that all RMC
models have essentially the same bond-angle distributions. Differences
lie within the statistical spread.

\begin{figure}[tbh]
\includegraphics[width=0.48\textwidth,clip=.]{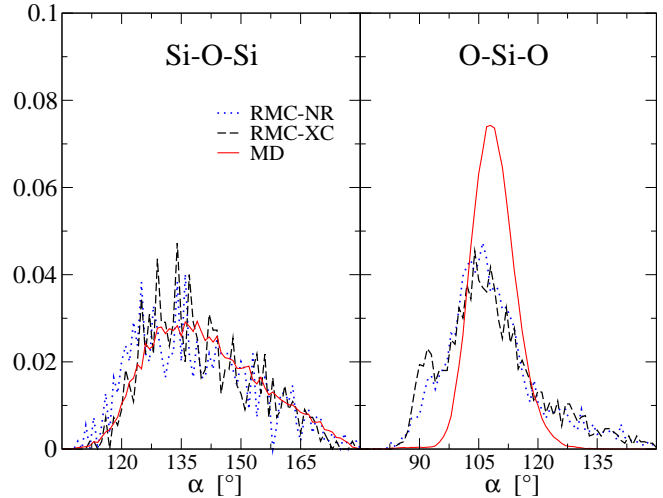}
\caption{\label{fig:angle} (color online) Bond-angle distributions of
the MD structure (red solid lines) and RMC models based on a
crystalline starting configuration using neutron diffraction data
(black dashed lines) as well as based on a random starting
configuration using X-ray diffraction data (blue dotted lines). In the
left panel the distributions of intra-tetrahedral angles (O-Si-O), and
in the right panel the distributions of inter-tetrahedral angles
(Si-O-Si) are shown. }
\end{figure}

In Fig.~\ref{fig:angle} the distributions for the RMC-NR and RMC-XC
model are compared to that of the MD structure. The Si-O-Si bond-angle
distribution of the RMC models agrees well with the MD data. The
intra-tetrahedral bond-angles, on the other hand, are much broader
distributed in the RMC models than in the MD structure. This
impression can be quantified by calculating the mean angles
$\bar\alpha(\mbox{Si-O-Si})$ and $\bar\alpha(\mbox{O-Si-O})$ as well
as the standard deviations $\Delta\alpha(\mbox{Si-O-Si})$ and
$\Delta\alpha(\mbox{O-Si-O})$. It is found that the mean angles of all
RMC models agree very well with the MD values, while the standard
deviations are larger by a factor of two, see Table~\ref{tab:angle}.
This finding corresponds to the deviations observed in the partial
radial distribution functions in Fig.~\ref{fig:prdf} and the
coordination number distribution of O-O in Fig.~\ref{fig:coord}.
There, distinctive features of the MD data, as the second peak in
$g^{\rm MD}_{\rm OO}$ and the bimodal distribution in $f^{\rm MD}_{\rm
OO}$, are not well reproduced by the RMC models.

\begin{table}
\caption{\label{tab:angle} {Mean values and standard deviations of bond angles.}}
\begin{footnotesize}
\begin{tabular}{cccccccccc}
\hline\hline
 &MD&XC&XR&NC&NR&NXR&IC&IR&C\\\hline
 $\bar\alpha(\mbox{Si-O-Si})$ 	& 141.2	& 139.8 & 140.7 & 137.4 & 138.6 & 140.5 & 138.0 & 138.5 & 125.7 \\
 $\Delta\alpha(\mbox{Si-O-Si})$ 	& 13.7	& 13.1 & 13.6 & 14.5 & 14.5 & 13.1 & 14.9 & 15.5 & 0.5\\
 $\bar\alpha(\mbox{O-Si-O})$ 	& 108.8	& 108.7 & 108.7 & 108.7 & 108.8 & 109.0 & 108.2 & 108.2 & 108.9\\
 $\Delta\alpha(\mbox{O-Si-O})$ 	& 5.9	& 13.1 & 12.9 & 12.7 & 12.4 & 11.3 & 15.6 & 15.9 & 3.2\\
\hline\hline
\end{tabular}
\end{footnotesize}
\end{table}

\subsection{\label{sec:rings} Ring-size distribution}

\begin{figure}[tbh]
\includegraphics[width=0.45\textwidth,clip=.]{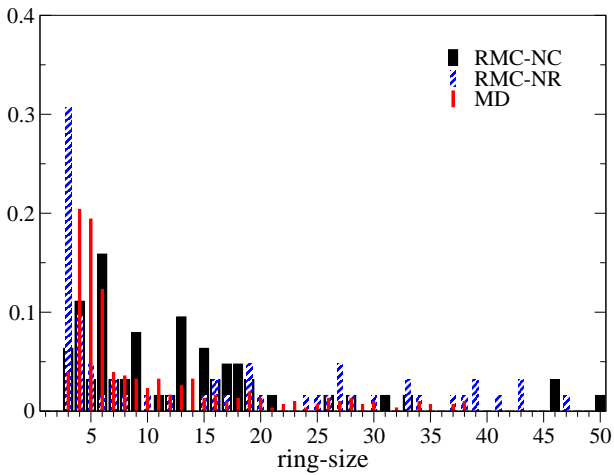}
\caption{\label{fig:rings} (color online) Comparison of ring-size distribution of
Li$_2$O-SiO$_2$ RMC models based on a crystalline starting
configuration using neutron diffraction data (black wide bars) and
based on a random starting configuration (blue striped bars) with that
of a Li$_2$O-SiO$_2$ MD system (red narrow bars).}
\end{figure}

In order to compare the topography of the glass-network we determined
ring-size distributions for each model. Here rings and their size are
defined in the following manner: 
\begin{itemize}
\item[(i)] A Si-atom and an O-atom are considered neighbors if their
distance is smaller than 2.0~{\AA} (using closest image convention).
\item[(ii)] For each Si-atom $\gamma$, the smallest closed loop of
alternating neighboring silicon and oxygen atoms is determined, which
entails the Si-atom $\gamma$.
\item[(iii)] The size of the ring is equal to the number of its
Si-atoms. 
\end{itemize}
The maximum number of rings equals the total number of
Si-atoms. However the number of rings is generally smaller, since
there are a number of Si-atoms for which no ring is found (e.~g., for
an isolated SiO$_4$ tetrahedra), and two different Si-atoms can be
associated with the same ring.

In Fig.~\ref{fig:rings} the ring-size distributions of the MD system,
the RMC-NC and the RMC-NR-model are shown. No data is shown for the
X-ray and combined data based RMC models, since their ring-size
distributions are practically the same as for the neutron diffraction
data based RMC models. Indeed the ring-size distributions almost do
not change compared to those of the initial models. On the other hand,
there is a clear dependence upon the starting configurations. While
most rings (40\%) of the RMC-NR model are of size three and four, the
RMC-NC model has a high number of rings of size 4 and 6. The latter is
more in line with what is found in the MD system.

\begin{figure}[tbh]
\includegraphics[width=0.45\textwidth,clip=.]{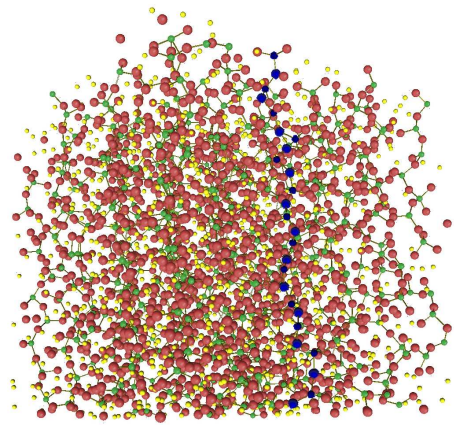}
\caption{\label{fig:chain} (color online) Picture of the RMC model
based on a crystalline starting configuration. Lithium atoms are
marked as small yellow spheres, Silicon atoms as medium sized green
spheres, and Oxygen atoms as large red spheres. In the center a linear
chain of neighboring Si and O is marked in dark blue, which is a
leftover of the starting configuration. }
\end{figure}

The RMC-NC model has a high number of rings of a size larger than 10.
Examining these large rings in more detail reveals that most of them
are actually straight linear chains penetrating the system parallel to
one system axis (they are seen as rings due to the periodic boundary
conditions, and can be found in both the RMC-NC and the RMC-XC models;
see also Fig.~\ref{fig:chain}). Such straight chains are not seen in
the RMC-NR and the RMC-XR models and in the MD structure. There are
also large rings and chains in these models, but those are generally
much more twisted than in the RMC models based on the crystalline
starting configurations.

\section{Summary and Conclusions}

The RMC method successfully reproduces many salient features of the
local structure of the original MD system. Some differences are found
in the partial radial distribution function of O-O, and in the
coordination number distributions of Si-Si and O-O. With respect to
the structure beyond nearest neighbor distances the RMC models are
less predictive and therefore cannot be expected to capture the medium
range order properly.

Comparing RMC models based on X-ray and neutron scattering data
revealed no significant differences. Moreover, we found that the
additional consideration of scattering data from one type of probe (either
X-ray or neutron) gives only a modest improvement over the initial 
RMC models that are based on geometric constraints only (number
density, minimal pair distances, some coordination numbers).
The situation becomes much better, however, when including the
information from both types of scattering probes.

Most structural properties of the RMC models do not depend sensitively
on the starting configuration (crystalline or random). Even the
ring-size distributions do not differ that much. However, taking a
closer look at the rings, revealed that the RMC models based on the
crystalline starting configuration exhibit straight linear chains
penetrating the system. These straight chains are remnants of the
crystalline starting configuration and their occurrence is not
reflected in the other structural properties studied. In particular,
there are no differences in $G^{\rm PDF}(r)$ and $S(Q)$ between the
RMC models based on the crystalline and the random starting
configuration. These findings show that one should check carefully if
a feature of interest in RMC models is only a product of a particular
starting configuration or if it can be reproduced using totally
different starting configurations.

\begin{acknowledgments}
Work on this project was supported in the Materials World Network by
the Deutsche Forschungsgemeinschaft (DFG Grant number MA~1636/3-1) and
by the NSF (NSF DMR Grant number 0710564).
\end{acknowledgments}


\begin{thebibliography}{10}
\bibitem{comm-random}
{This starting configuration does not correspond to a fully random arrangements
  of the ions, but to a quasi-random sequential setup as described in
  Sec.~9.8.2. of the manual of the RMCA program, see
  Ref.~\onlinecite{rmc++-man}.}
\end{thebibliography}

\end{document}